\documentclass[journal]{IEEEtran}

\usepackage{setspace}
\usepackage{cite,graphicx,amsmath,amssymb}
\usepackage{subfigure}
\usepackage{url}
\usepackage{fancyhdr}
\usepackage{mdwmath}
\usepackage{mdwtab}
\usepackage{balance}
\usepackage{xcolor}
\usepackage{bm}
\usepackage{amsthm}
\usepackage{threeparttable}
\usepackage{algorithm}
\usepackage{algorithmic}
\usepackage{multirow}
\usepackage{flafter}
\usepackage{makecell}
\usepackage{diagbox}
\usepackage{booktabs}
\usepackage{multirow}
\usepackage{siunitx}

\usepackage{comment}

\usepackage{float}
\floatstyle{plaintop}
\restylefloat{table}
\usepackage{caption}

\providecommand{\url}[1]{#1}
\allowdisplaybreaks
\begin{document}
\title{Pinching-Antenna Systems (PASS): Architecture Designs, Opportunities, and Outlook}
\author{Yuanwei Liu, Zhaolin Wang, Xidong Mu, Chongjun Ouyang, Xiaoxia Xu, and Zhiguo Ding\\
\thanks{Yuanwei Liu is with The University of Hong Kong and also with Kyung Hee University. Zhaolin Wang is with The University of Hong Kong. Chongjun Ouyang and Xiaoxia Xu are Queen Mary University of London. Xidong Mu is with Queen's University Belfast. Zhiguo Ding is with Khalifa University and also with The University of Manchester.}
\vspace{-0.5cm}
}

\maketitle
\begin{abstract}

Flexible-antenna systems have recently attracted significant research attention due to their potential to intelligently reconfigure wireless channels. However, the current flexible-antenna systems still suffer from fundamental limitations, such as free-space path loss and line-of-sight blockage. This article introduces a novel flexible-antenna system, termed the Pinching-Antenna SyStem (PASS). PASS adopts the dielectric waveguides as the primary transmission medium and radiates signals into free space by flexibly pinching discrete dielectric particles, referred to as pinching antennas, along the waveguide. By combining the strengths of both wireless and wired communication, PASS effectively mitigates inherent wireless limitations while offering high antenna reconfigurability. This article reviews the key features of PASS in comparison with conventional wireless systems, analyzes its main advantages, and discusses potential designs, transmission architectures, and application scenarios. Finally, it outlines promising research directions and open challenges associated with PASS.


\end{abstract}

\section{Introduction}
The perspective introduced by Shannon's information theory established the foundation for the digital revolution and has profoundly shaped the development of modern communication systems \cite{shannon1948mathematical}. In Shannon's original framework, the channel connecting the transmitter and receiver was treated as a fixed entity, i.e., a parameter entirely dictated by environmental factors and beyond human control. However, advances in communication technology have challenged this assumption by enabling active reconfiguration of channel parameters to boost the channel throughput. A notable example is the deployment of multiple-input multiple-output (MIMO) systems, which reconfigures the channel by creating separated parallel point-to-point links to enhance communication efficiency \cite{bjornson2024introduction}.

Building on this momentum, recent advancements in intelligent and flexible antenna designs have further introduced novel approaches for channel reconfiguration. Technologies, such as reconfigurable intelligent surfaces \cite{9140329,8910627}, dynamic metasurface antennas \cite{shlezinger2021dynamic}, fluid antennas \cite{10753482}, and movable antennas \cite{10286328}, can dynamically adjust their positions, apertures, and electromagnetic properties, enabling precise control of wireless channel conditions. By optimizing their reconfigurable parameters, these novel antenna types and array architectures offer flexibility in controlling the effective end-to-end channel gain. This capability has demonstrated great potential for improving system performance and has attracted considerable research interest.

Despite their potential to improve network throughput, current flexible-antenna systems, in which antenna reconfigurations are often confined to small apertures, face two fundamental wireless limitations: free-space path loss and line-of-sight (Los) blockage. Free-space path loss leads to rapid signal attenuation, particularly over long distances or at high frequencies, while LoS blockage can cause deep fading and create coverage dead zones. To overcome these challenges, a novel, highly flexible, and efficient design, known as the Pinching-Antenna SyStem (PASS) \cite{suzuki2022pinching,ding2024flexible}, has been recently proposed.


\begin{figure*}[t!]
  \centering
  \includegraphics[width=0.8\textwidth]{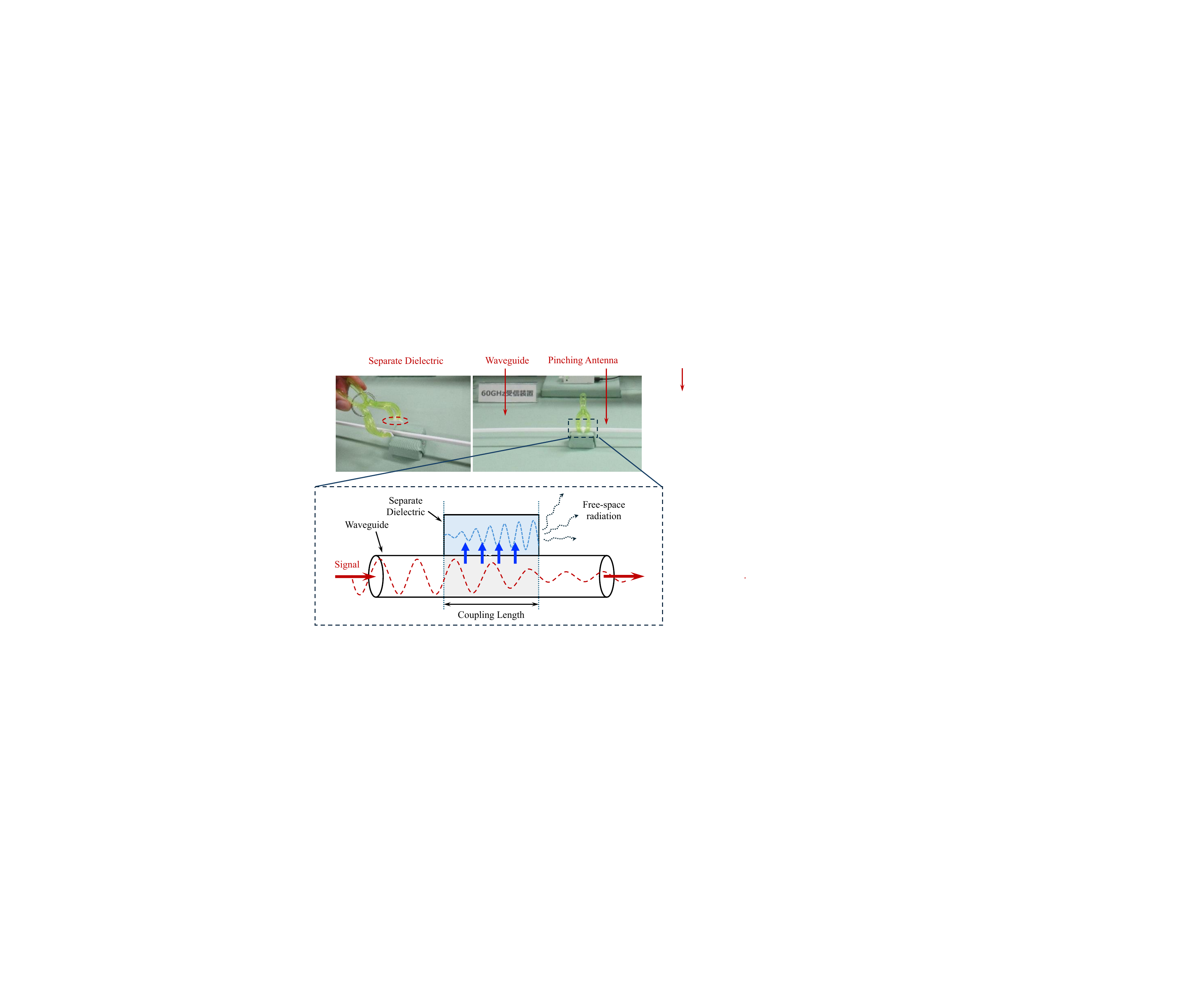}
  \caption{Comparison between PASS and conventional MIMO systems, along with an illustration of the physical principles underlying PAs and NTT DOCOMO's prototype (Photo credit: NTT DOCOMO) {\copyright} CCBY \cite{suzuki2022pinching}.}
  \label{fig_waveguide_model}
  \vspace{-0.3cm}
\end{figure*} 

\subsection{The Road to Near-Wired Connectivity: A Prototype of PASS}

The most effective way to overcome free-space path loss and LoS blockage is through cable-based wired communication, which completely eliminates both issues. However, wired links are inherently unsuitable for mobile scenarios. This limitation raises a fundamental question: \emph{Can we bridge the gap between fixed cables and fully over-the-air communication?} A promising answer is PASS, first developed and prototyped by NTT DOCOMO in 2021 \cite{suzuki2022pinching}, as illustrated in Fig. \ref{fig_waveguide_model}. PASS is based on low-loss dielectric waveguides that serve both as a transmission medium and as an “antenna rail.” At an arbitrary point along the waveguide, a small dielectric particle can be pinched onto the waveguide, like attaching a clothespin to a line, to form a \emph{pinching antenna} (PA). Each PA functions as a controlled leakage point, enabling the guided wave to radiate into or receive energy from free space. This makes PASS act as a reconfigurable waveguide and provides greater flexibility than traditional leaky-wave antennas.

Unlike conventional flexible-antenna systems, a PASS waveguide can be run for tens of meters or even longer, allowing PAs to be positioned within LoS of users \cite{suzuki2022pinching, ding2024flexible}. This adaptability effectively creates a \emph{near-wired} connection, ensuring favorable channel conditions and shifting the focus of wireless communication networks from the \emph{“last mile”} to the more manageable \emph{“last meter”}. Therefore, PASS combines the high reliability of wired communication with the deployment flexibility of wireless access, offering a novel and effective approach to overcoming the limitations of conventional wireless systems. This unique capability is reflected in the name “PASS”, highlighting its ability to provide the shortest and most efficient wireless connections between the PAs and the wireless transceivers, which effectively \emph{pass} wireless signals with minimal loss.

\subsection{Key Advantages for Employing PASS in Wireless Communications}
Considering the above unique features, the key advantages of PASS are summarized as follows:
\begin{itemize}
\item \textbf{Enhanced LoS Stability and ``Last Meter'' Communications:} By extending the waveguide and enabling the flexible deployment of PAs, PASS effectively mitigate large-scale path loss and establishes stable LoS communication environments. This capability enhances \emph{``last meter'' communications} and expands network coverage, which is not achievable with other flexible-antenna technologies.
\item \textbf{Flexible Pinching Beamforming Gain:} PASS control signals by optimizing the locations and the numbers of PAs, which enables highly flexible and scalable \emph{pinching beamforming}. This capability allows for precise management of signal strength and interference, effectively meeting stringent service requirements.
\item \textbf{Utilization of Near-Field Benefits:} The extended aperture of the waveguide places users within the near-field region of PASS, which enables the use of near-field electromagnetic properties, such as beam focusing, to harvest the spatial degrees of freedom and enhance wireless transmission efficiency \cite{liu2024near}.
\item \textbf{Cost-Effective and Scalable Implementation:} PASS are simple to implement and economical to deploy, not only because of the low cost of waveguides, but also because PASS only require adding or removing dielectric materials at selected points along the waveguide \cite{suzuki2022pinching}. Furthermore, PASS theoretically allows the addition or removal of any number of radiating units along the waveguide in a low-cost manner, making them a scalable and flexible solution for practical network design.
\end{itemize}
These advantages have fueled growing research interest in PASS. The performance of PASS has been analyzed in terms of its ergodic rate \cite{ding2024flexible} and its array gain \cite{ouyang2025array}. Furthermore, various activation algorithms have been proposed to optimize the locations of PAs along a single waveguide, such as \cite{ding2024flexible} and \cite{wang2025modeling}. However, despite these initial efforts, the design of PASS for wireless communications remains in its early stage. Therefore, this article aims to provide a concise yet comprehensive overview of existing PASS techniques and to offer new insights into their development, serving as a useful reference for researchers interested in exploring the vast potential of PASS. We begin by discussing the fundamental differences between PASS and conventional multi-antenna systems. Next, we introduce several potential transmission architectures to support the design of pinching beamforming. We then explore promising application scenarios and evaluate the corresponding system performance. Finally, we outline key open challenges to stimulate further research in this emerging area.

\section{Pinching Antenna Fundamentals:\\What Will be Different?}

In this section, we introduce the fundamentals of PAs and highlight the basic differences from conventional electronic antennas from their physical principle, signal model, and communication system design perspective, respectively. 

\subsection{Physical Principle}
Unlike conventional electronic antennas, which rely on electronic devices to convert alternating electric currents into electromagnetic (EM) waves, the physical principle of PAs is based on the phenomenon of EM coupling that arises when a separate dielectric is brought into close proximity with a dielectric waveguide \cite{suzuki2022pinching}. This coupling enables the power exchange between the waveguide and the separate dielectric, as illustrated in the bottom-right of Fig. \ref{fig_waveguide_model}. Therefore, by pinching a separate dielectric onto the waveguide, the EM waves originally confined within the waveguide are coupled into the separate dielectric, where they are subsequently radiated into free space, thereby forming an “antenna”. Unlike conventional electronic antennas, PAs are entirely passive, simpler to manufacture and deploy, and hence provide greater position flexibility.

By modeling the pinching antenna as an open-ended directional coupler, the power exchange caused by the EM coupling can be characterized by the coupled-mode theory. Let $P_{\mathrm{guide}}$ and $P_{\mathrm{pinch}}$ denote the normalized power in the waveguide and the pinched separate dielectric, respectively. Their relationship is approximately given by \cite{okamoto2010fundamentals}
\begin{align}\label{power_relation}
    P_{\mathrm{guide}} &= 1 - F \sin^2(\kappa L), \quad P_{\mathrm{pinch}} = F \sin^2(\kappa L),
\end{align}
where $L$ is the coupling length shown in Fig. \ref{fig_waveguide_model}, $\kappa$ is the coupling coefficient, and $F \le 1$ is the maximum coupling efficiency. 
 When the waveguide and the separate dielectric share the same effective refractive index, the coupling efficiency $F$ reaches its maximum value of $1$, making it possible to radiate full power from a single PA with a coupling length of $\pi/(2\kappa)$. 
Moreover, from the above formula, it can be observed that the power emitted by each PA can be controlled by adjusting the coupling length $L$ and the coupling coefficient $\kappa$.
The coefficients $\kappa$ and $F$ are influenced by various factors, including the effective refractive index, transverse modes, and the spacing between the waveguide and the separate dielectric.

\subsection{Signal Model}
As shown in Fig. \ref{fig_waveguide_model}, the composite channel from the base station (BS) to the user in the PASS system consists of two components: in-waveguide propagation and free-space propagation. Compared to free-space propagation, in-waveguide propagation incurs almost negligible path loss. For example, a ceramic-ribbon dielectric waveguide operating in the millimeter-submillimeter band attenuates by less than 0.01 dB/m \cite{yeh2000communication}. Consequently, when multiple PAs are deployed on a waveguide, the signal received at the user from the $n$-th PA through the LoS channel can be modeled as:
\begin{equation}
    y_n = \frac{\beta_n \sqrt{P_n}}{r_n} e^{-j \frac{2 \pi}{\lambda} \left( r_n + n_{\mathrm{eff}} d_n \right)} x.
\end{equation}
In this expression, $j = \sqrt{-1}$, $x$ is the signal fed into the waveguide, $\beta_n$ is the complex channel of the free-space propagation at the reference distance of 1 meter, $P_n$ is the power radiated from the PA, $\lambda$ denotes the signal wavelength, $n_{\mathrm{eff}}$ represents the effective refractive index of the waveguide, $r_n$ is the distance of the free-space propagation from the PA to the user, and $d_n$ is the distance of the waveguide propagation. By adjusting the location of PAs, both $r_n$ and $d_n$ can be modified, which in turn affects the amplitude and phase of the received signal.


Based on the physical principle of PAs, the power model of PASS, signified by $P_n$, is significantly different from that of conventional multi-antenna systems. Specifically, with conventional antennas, the power of the signal emitted from each antenna is actively controlled by electronic devices, including radio-frequency (RF) chains and power splitters. In contrast, for PAs, the transmit power is passively determined by the coupling length with the waveguide. Furthermore, when multiple PAs are pinched along a waveguide, the power radiated by subsequent antennas is influenced by the power radiated by preceding ones, leading to a coupled power model. While the coupling length of a PA is not as easily adjustable as the electronic parameters of conventional antennas, since PAs are typically passive, it is still possible to design and manufacture PAs with specific coupling lengths to achieve a desired power model. In the following, we introduce two common power models for PAs.
\begin{itemize}
    \item \textbf{Equal Power Model:} In this model, the PAs on the waveguide radiate an equal amount of power. This can be achieved by designing the PAs with varying coupling lengths, such that the earlier PAs radiate a smaller proportion of the power in the waveguide, while the later PAs radiate a larger proportion of the remaining power.
    \item \textbf{Proportional Power Model:}  In this model, each PA is designed with an identical coupling length, thereby radiating the same fixed proportion of the power left in the waveguide. As a result, the power radiated by each PA decreases progressively along the waveguide, where the power radiated by later PAs is proportional to that of earlier ones by a fixed ratio. 
\end{itemize}
These two power models offer different pros and cons. Compared to the proportional power model, where the latter PAs may radiate only a smaller amount of power, the equal power model ensures that all PAs are effectively utilized. The equal power model also simplifies the performance analysis of the PASS system due to its simple mathematical formulation, which is useful to identify the fundamental limits of PASS, and therefore has contributed to its widespread adoption in the literature. In contrast, the proportional power model requires less stringent manufacturing specifications, as all PAs share the same design, thus reducing production complexity and costs.

\begin{figure}[t!]
  \centering
  \includegraphics[width=0.4\textwidth]{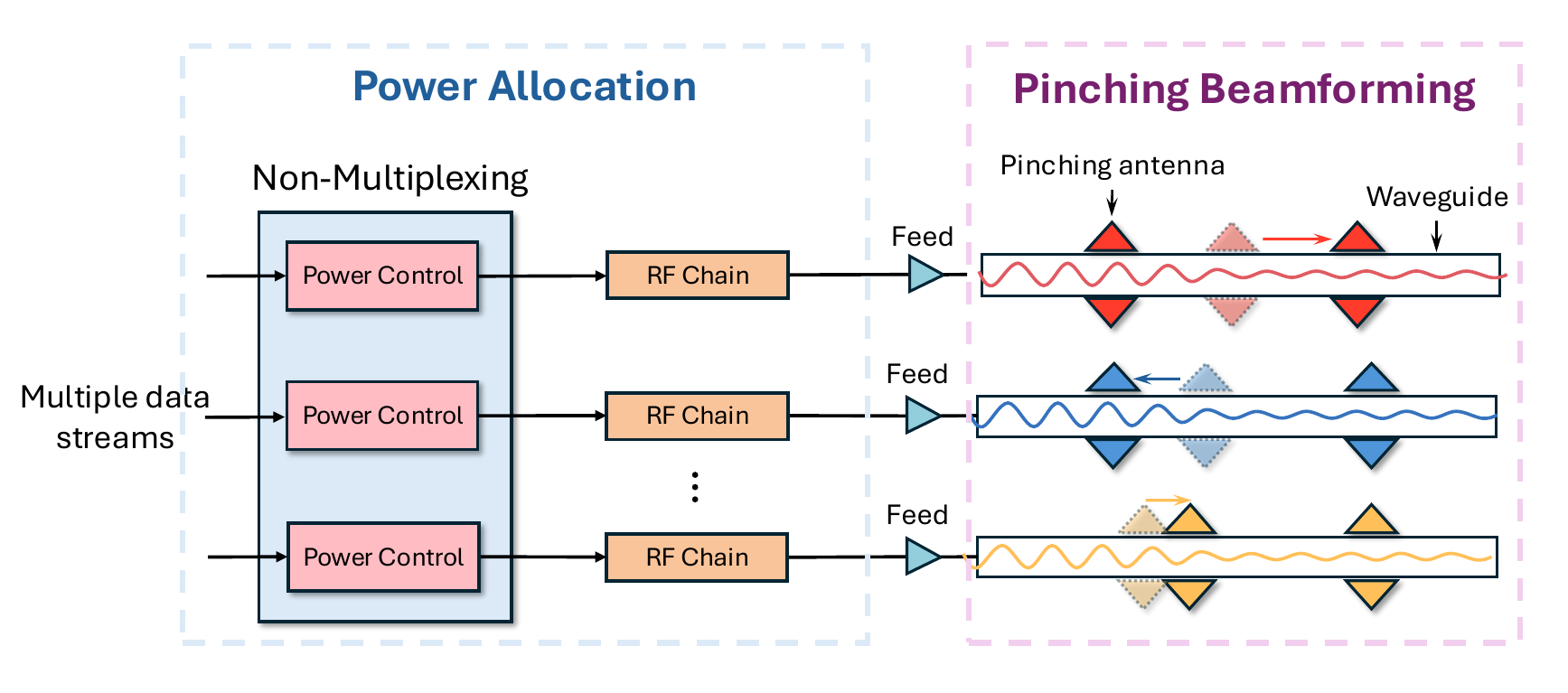}
  \caption{Waveguide division architectures for PASS}
  \label{fig_non_multiplex_architecture}
\end{figure} 

\subsection{Communication System Design}
From a communication system design perspective, it is crucial to position the PAs at suitable locations to maximize communication performance. On the one hand, the PAs should be deployed close to the user to mitigate the significant signal attenuation caused by free-space path loss and blockage. On the other hand, when multiple PAs are deployed along the waveguide, they form an antenna array, enabling \emph{pinching beamforming} to enhance signal strength and reduce interference. Note that pinching beamforming poses greater challenges compared to existing flexible-antenna systems. This is because fine-tuning the positions of the PAs along the waveguide enables simultaneous adjustment of the phase shifts of in-waveguide and free-space propagation, as well as the free-space path loss, to achieve the desired beamforming gain. Leveraging the flexibility and portability of PAs, their positions can be determined either actively by the BS or passively by the user, leading to the following two designs:
\begin{itemize}
    \item \textbf{BS-centric Design:} In this design, the positions of PAs are determined by the BS to optimize specific performance metrics of the overall PASS system, such as spectral efficiency and user fairness. In dynamic scenarios, there are two typical PA activation schemes, namely \emph{continuous activation} and \emph{discrete activation}. For continuous activation, a slide track is installed parallel to the waveguide, allowing the attached PAs to be mechanically repositioned along the waveguide, e.g., using micro motors. For discrete activation, a large number of PAs are pre-installed at various locations along the waveguide and dynamically activated through fast-response electrical means, such as using electromagnetic relays. It is worth noting that mechanical continuous activation may introduce considerable time delays, making it less efficient for fast-varying wireless channels compared to electrical discrete activation. However, it offers more precise PA positioning and is preferable for slow-varying wireless environments.
    \item \textbf{User-centric Design:} In this design, users can pinch or release portable PAs onto the waveguide as needed to enhance communication service in their local area. Whenever a PA is pinched or released by a user, the baseband processing strategy needs to be optimized and updated based on the new propagation environment to avoid performance degradation for other users.
\end{itemize}
By comparing these two designs, it can be concluded that the BS-centric design offers high flexibility, allowing the PASS to effectively mitigate path loss and blockage issues while enhancing beamforming gain for all users. However, this comes with a higher implementation cost. In contrast, the user-centric design has a lower implementation cost but may result in each PA being effectively utilized for a single user.

\vspace{-0.1cm}
\section{Transmission Architectures for PASS}

In this section, we discuss the viable transmission architectures for PASS, while identifying their respective advantages and disadvantages.
 
\vspace{-0.1cm}
\subsection{Waveguide Division Architecture}
A schematic illustration of the waveguide division architecture is shown in Fig. \ref{fig_non_multiplex_architecture}. In this architecture, the data streams to be transmitted are not multiplexed at the baseband, meaning that each waveguide is fed individually by a single data stream. In this setup, the primary role of the baseband processing is to manage the power allocation across these data streams to optimize transmission efficiency. The key benefit of this design is to reduce implementation complexity. Compared to the conventional multi-antenna systems, the unique capability of pinching beamforming, achieved by adjusting the positions of PAs, can be utilized to enhance communication performance. This waveguide division architecture is particularly well-suited for those commonly encountered scenarios, where each waveguide serves geographically isolated areas, minimizing the need for advanced interference management between waveguides. 

\begin{figure}[t!]
  \centering
  \includegraphics[width=0.4\textwidth]{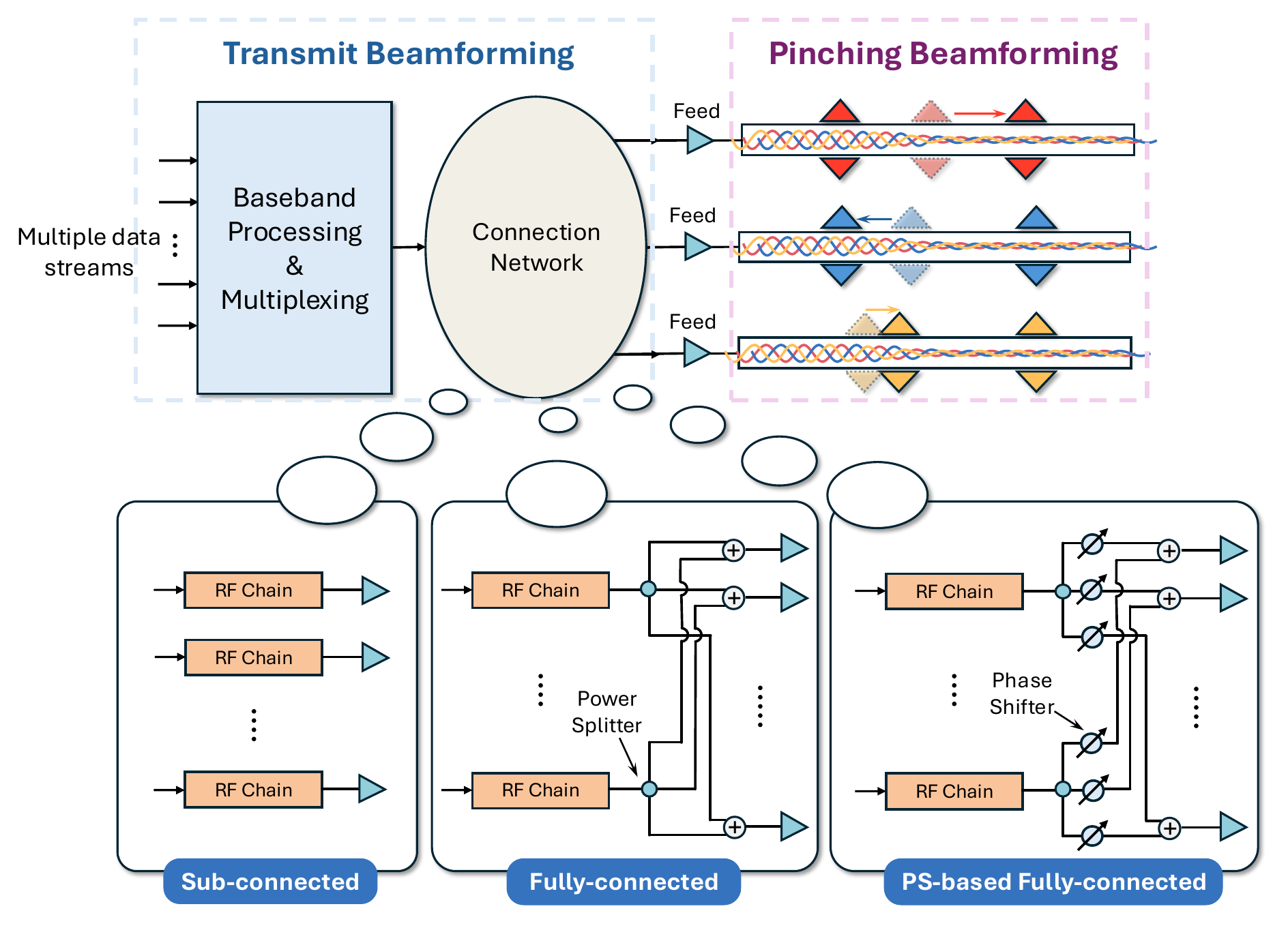}
  \caption{Waveguide multiplexing architectures for PASS}
  \label{fig_multiplex_architecture}
\end{figure}

\begin{figure*}[t!]
  \centering
  \includegraphics[width=0.8\textwidth]{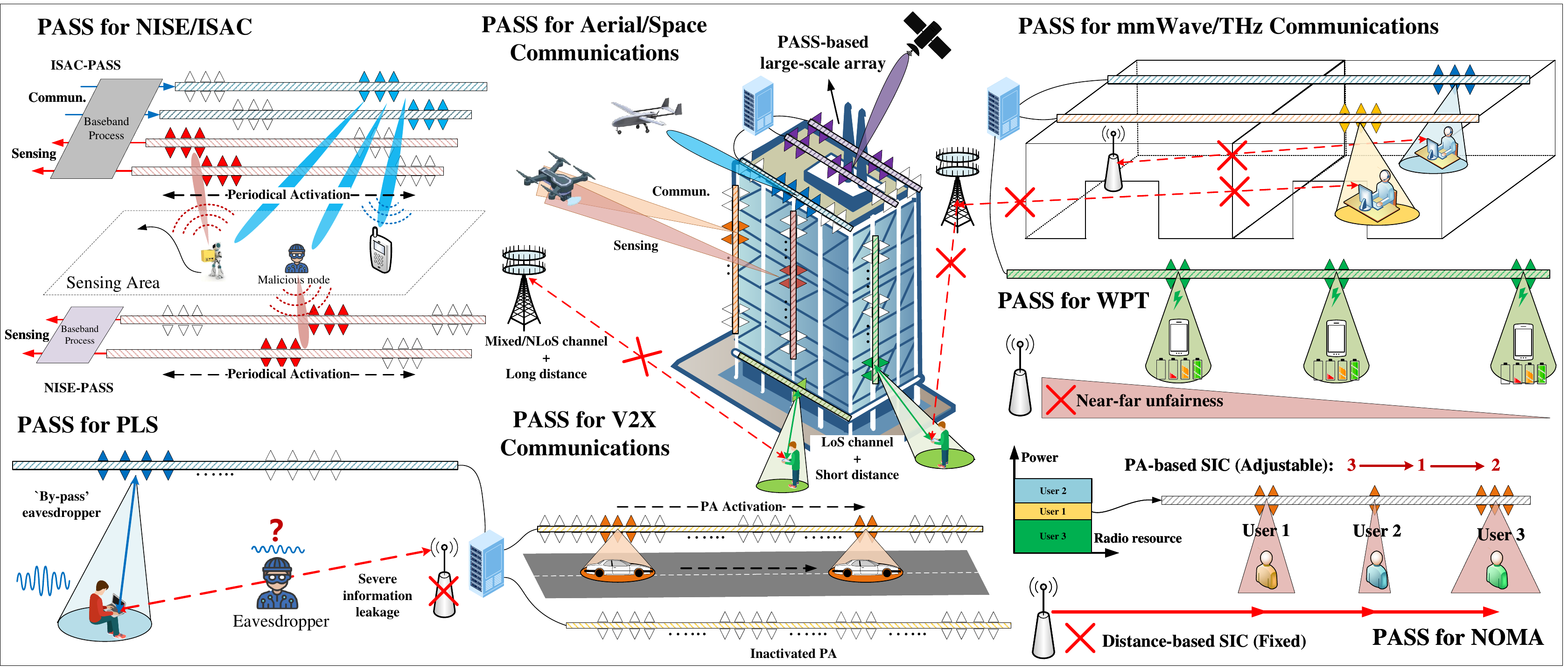}
  \caption{Outdoor/Indoor deployments and emerging applications of PASS in 6G and beyond.}
  \label{fig_application}
  \vspace{-0.3cm}
\end{figure*} 

\vspace{-0.1cm}
\subsection{Waveguide Multiplexing Architecture}
As shown at the bottom of Fig. \ref{fig_multiplex_architecture}, the waveguide multiplexing architecture multiplexes multiple data streams before feeding them into the waveguides, thus facilitating the \emph{joint transmit and pinching beamforming}. This architecture can be seen as an extension of the conventional MIMO transmission architecture, where each electronic antenna is replaced by a waveguide. The key difference between the two architectures lies in the point of signal radiation. In conventional MIMO systems, signals are radiated directly from the electronic antennas into free space. In contrast, in the waveguide multiplexing architecture of PASS, signals are radiated at specific locations along the waveguide via PAs. Consequently, unlike conventional MIMO, which relies on a massive number of electronic antennas to achieve high spatial degrees of freedom (DoFs) \cite{6736761}, the waveguide multiplexing architecture significantly reduces the requirement for numerous waveguides because a large number of low-cost PAs can be easily installed on a single waveguide. Based on the above discussion, the waveguide multiplexing architecture can be designed in analogy to conventional MIMO architecture, but with distinct differences in implementation and design focus. In the following, we introduce three basic waveguide multiplexing architectures, as shown in Fig. \ref{fig_multiplex_architecture}, namely \emph{sub-connected architecture}, \emph{fully-connected architecture}, and \emph{phase shifter (PS)-based fully-connected architecture}.
\begin{itemize}
    \item \textbf{Sub-connected Architecture:} This is the simplest multiplexing architecture, where each waveguide is connected to a dedicated RF chain and carries signals that are multiplexed and precoded by baseband digital beamforming. While this architecture is analogous to the fully-digital architecture used in conventional MIMO systems, it is referred to as “sub-connected” because the signals from each RF chain are emitted exclusively by those PAs deployed on the corresponding waveguide. 
    \item \textbf{Fully-connected Architecture:} In this architecture, the signals from each RF chain are fed into all waveguides with the aid of power splitters. The fully-connected architecture addresses two key limitations of the sub-connected design. First, it enables each RF chain to connect to all PAs across all waveguides, thereby fully using the overall spatial DoFs to enhance communication performance. Second, the number of RF chains no longer needs to match the number of waveguides. This flexibility allows for either reducing the number of RF chains to save energy, as in hybrid beamforming for millimeter-wave networks, or increasing it to further exploit the additional spatial DoFs provided by the massive PAs.
    \item \textbf{PS-based Fully-connected Architecture:} This architecture builds on the fully-connected design by introducing phase shifters (PSs) to configure the signals before feeding them into the waveguides. This enhancement enables the realization of \emph{tri-hybrid beamforming}, which integrates baseband digital beamforming, analog beamforming enabled by the PSs, and pinching beamforming facilitated by the PAs, exhibiting the potential to significantly improve the performance of PASS.
\end{itemize}




\section{“PASS” Signals for 6G and Beyond}

In this section, we will discuss promising applications of PASS to underpin 6G and beyond.

\vspace{-0.1cm}
\subsection{PASS for Ubiquitous Indoor and Outdoor Coverage}
The low-complexity structure of PASS enables it to be easily deployed on diverse structures in both outdoor and indoor scenarios, such as building facades, rooftops, roadside guardrails, and indoor ceilings or walls, as depicted in Fig. \ref{fig_application}. PASS addresses the key challenge of conventional wireless communication, where communication links are often partially or totally blocked by obstacles and suffer from significant long-distance path loss due to the remotely deployed BSs. In particular, PASS establishes \emph{LoS-dominated} and \emph{short-distance} wireless channels with the widespread deployment of waveguides and the optimal activation of PAs near the terminals. By preventing wireless signals from being wasted or even interfering with other users, PASS is capable of facilitating ubiquitous outdoor and indoor coverage in a cost-efficient and energy-efficient manner. Several attractive applications of PASS in indoor and outdoor scenarios will be discussed below.

On the one hand, PASS is quite appealing in indoor scenarios. As shown in the top right of Fig. \ref{fig_application}, PASS installed on the ceiling can flexibly activate PAs to provide wireless communications for users in different rooms. This mitigates the severe penetration losses suffered by wireless signals transmitted by outdoor BSs or indoor access points (APs), particularly at mmWave and THz frequencies. Moreover, by exploiting the near-field channel between PAs and users, efficient beam focusing can be achieved. This not only enhances the desired signal strength but also reduces interference to other users. On the other hand, the widely distributed PASS can also be utilized in outdoor mobile scenarios, such as aerial/space communications and vehicular communications, as illustrated in the middle of Fig. \ref{fig_application}. For example, recall the fact that conventional BSs with down-tilt antennas are typically inefficient for aerial communications. As a remedy, PASS installed on rooftops and facades can realize tailored upward beamforming for communications and sensing to effectively support medium- and high-altitude aircrafts as well as low-altitude unmanned aerial vehicles (UAVs) or other high-altitude platforms. As a further advancement, PASS can be employed in space communications. By integrating multiple waveguides and numerous PAs, PASS can form a large-scale antenna array to combat the severe path loss typically encountered in high-frequency satellite communications. Moreover, with PASS uniformly deployed along roadside guardrails, efficient vehicular communications can be achieved by adaptively activating PAs matching with vehicular movement. This is particularly beneficial for autonomous driving in future wireless networks. 

\begin{figure}[t!]
  \centering
  \includegraphics[width=0.4\textwidth]{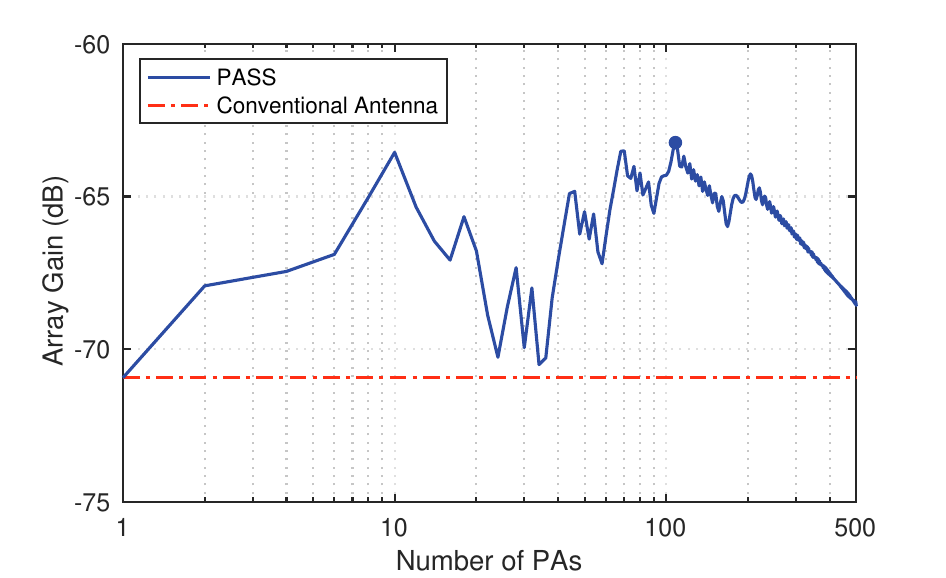}
  \caption{Array gain versus the number of pinching antennas. The detailed simulation setup is given in \cite{ouyang2025array}.}
  \label{fig_array_gain}
\end{figure} 

\vspace{-0.1cm}
\subsection{Interplay between PASS and NOMA}
The combination of PASS and non-orthogonal multiple access (NOMA) provides a “win-win” solution for 6G and beyond. On the one hand, since multiple PAs on a specific waveguide can only be fed with the same set of signals, how to serve multiple users simultaneously becomes a critical challenge. To address this, it is natural to use NOMA to multiplex multiple users' signals in the power domain over the same radio resource~\cite{ding2024flexible}. On the other hand, PASS also facilitates flexible NOMA operations. Recall the fact that NOMA relies on successive interference cancellation (SIC) to handle co-channel interference, and therefore SIC order significantly affects the communication performance achieved by each NOMA user. Conventionally, a fixed distance-based SIC order is employed for NOMA, where users closer to the BS, so-called strong users, are assigned higher SIC orders due to better channel conditions. In contrast, as shown in the bottom right of Fig. \ref{fig_application}, PASS-based NOMA enables a more flexible SIC operation. This is achieved by dynamically adjusting users' channel conditions through the allocation of PAs and the design of their activation locations. For example, users originally farther from the BS/AP can have higher channel conditions with more numbers of PAs allocated. By doing so, more customized wireless services can be provided by the PASS-based NOMA.

\vspace{-0.15cm}
\subsection{PASS for NISE/ISAC}
In addition to enhancing communication, PASS significantly improves sensing quality and coverage. Since the links between PAs and sensing targets follow the near-field model, PASS enables near-field sensing (NISE) to precisely estimate angular and range information, as well as target velocity. As illustrated in the top left of Fig. \ref{fig_application}, two practical structures can be employed: ISAC-PASS and NISE-PASS. (1) ISAC-PASS allocates part of the waveguides to transmit communication signals while using the rest to receive echo-sensing signals, enabling simultaneous near-field communication and sensing within the area of interest. (2) NISE-PASS, equipped exclusively with receiving waveguides, operates in a low-power mode to facilitate NISE. By leveraging the widespread deployment of PASS and successively or periodically activating sensing PAs, ubiquitous and highly accurate NISE can be achieved. This capability is particularly beneficial for real-world applications, such as tracking target trajectories or monitoring malicious entities.

\begin{figure}[t!]
  \centering
  \includegraphics[width=0.4\textwidth]{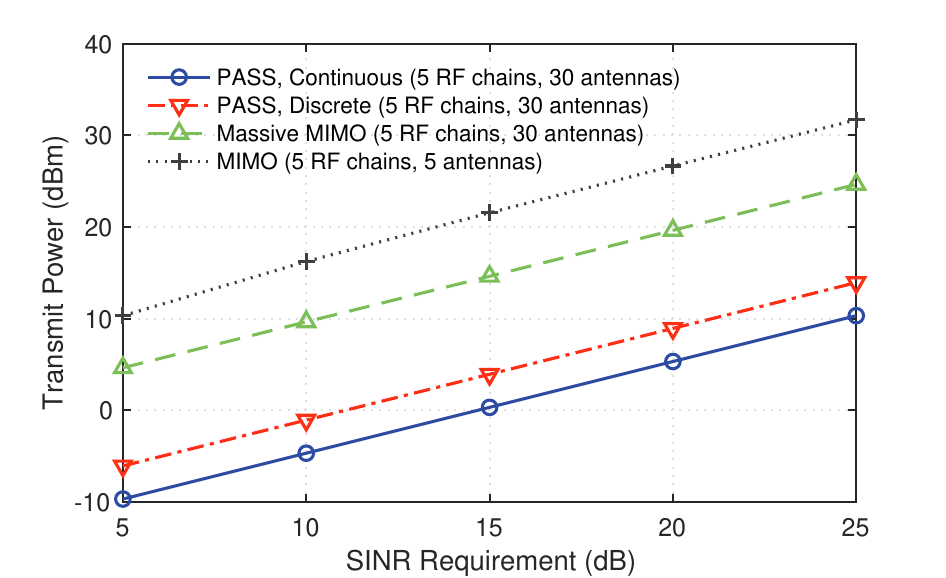}
  \caption{Minimum transmit power versus the minimum SINR. The detailed simulation setup is given in \cite{wang2025modeling}.}
  \label{fig_transmit_power_min}
\end{figure} 

\vspace{-0.15cm}
\subsection{PASS for WPT}
Wireless power transfer (WPT) is an efficient technology that uses RF signals to supply power to energy-constrained devices and extend their operational lifetime, especially for Internet-of-Things (IoT) and sensor networks. However, conventional WPT is hindered by the challenge of near-far unfairness, where devices located farther from the wireless power-supply BS or AP experience reduced energy-harvesting efficiency due to degraded channel conditions.  As shown in the middle right of Fig. \ref{fig_application}, this issue can be resolved through PASS-empowered WPT. In particular, tailored PAs can be activated right to each device, which ensures high-quality WPT for all devices, regardless of their locations.

\vspace{-0.15cm}
\subsection{PASS for PLS}
PASS can also enhance secure information transmission in the context of physical layer security (PLS). As shown in the bottom left of Fig. \ref{fig_application}, PASS enables legitimate information transmission to effectively bypass the eavesdropper. On the one hand, since signal transmission predominantly occurs within the waveguide of PASS, the likelihood of legitimate information being eavesdropped is significantly reduced compared to broadcast wireless channels. On the other hand, for the short-distance PA-user transmission, the near-field beam focusing can be employed to minimize the information leakage.

\vspace{-0.2cm}
\section{Numerical Case Studies}
In this section, we present numerical examples to characterize the performance of PASS employing pinching beamforming and to compare it with the conventional baselines.

\vspace{-0.2cm}
\subsection{Array Gains}
The array gain of PASS is analyzed by considering a setup where multiple PAs, equally spaced along a single waveguide, are used to serve a single-antenna user; see \cite[{\figurename} 1]{ouyang2025array}. This setup represents the most fundamental pinching beamforming scenario, and the corresponding results serve as a lower bound for those more advanced designs.
Unlike conventional fixed-location antenna systems, the wireless propagation in PASS is LoS-dominated and near-field-driven. Considering these factors, the array gain of the system has been analyzed with respect to the number of PAs, as illustrated in {\figurename} {\ref{fig_array_gain}} \cite{ouyang2025array}. The findings in {\figurename} {\ref{fig_array_gain}} indicate that the array gain of pinching antennas does not increase monotonically with the number of antennas. Instead, there exists an optimal number of PAs that maximizes the array gain achievable by PASS. Notably, the maximum array gain exceeds that of conventional antenna systems. This considered naive pinching beamforming represents a lower bound for other potential designs, whose observations lead to the following conclusions: \romannumeral1) Due to their flexibility, PASS can achieve a higher array gain than conventional fixed-location antenna systems. \romannumeral2) Optimizing the number of PAs is essential for PASS to fully realize its performance potential.

\vspace{-0.2cm}
\subsection{Signal Power Enhancement}

In Fig. \ref{fig_transmit_power_min}, we illustrate the effectiveness of PASS in enhancing signal power through the joint optimization of pinching beamforming and transmit beamforming. The penalty-based method is exploited to deal with this coupled optimization problem.Specifically, we consider a sub-connected architecture for PASS under a practical proportional power model. Both continuous activation and discrete activation are evaluated, where the latter incorporates candidate PAs to be pre-installed per half meter. For benchmarking, we employ two baseline schemes: a conventional MIMO system, where each RF chain connects to a single antenna, and a massive MIMO system, where each RF chain connects to multiple antennas via analog beamforming devices. All systems are configured with an identical number of RF chains, and only LoS channels are considered to enable a fair and equitable comparison. The results indicate that PASS significantly reduces the transmit power required to satisfy the signal-to-interference-plus-noise ratio (SINR) demands of communication users. Additionally, although discrete activation slightly limits beamforming flexibility compared to continuous activation, PASS still demonstrates substantial performance benefits. This improvement primarily stems from the establishment of robust LoS channels, effectively mitigating free-space path loss.

\section{Conclusions and Open challenges}
In this article, the promising concept of PASS was introduced, highlighting its great potential for enabling efficient wireless communications over the \emph{“last meter”}. The basic physical principle, signal model, and communication design of PASS have been discussed. By leveraging the novel pinching beamforming, two practical transmission architectures were developed for PASS, whose advantages and limitations were also described. Moreover, a range of promising applications for employing PASS in wireless networks were explored. Numerical case studies demonstrated the superiority of PASS over conventional multi-antenna transmission. As PASS remains in its early stages of research, there are numerous open challenges and opportunities to unlock its full potential. Several of these key research directions are exemplified below:
\begin{itemize}
    \item \textbf{Spatial Modeling for PASS with Stochastic Geometry}: Stochastic geometry provides a powerful framework for evaluating the average performance of wireless networks, which can be used to determine key design metrics, such as the required number of PAs, the ratio between PAs and users, and optimal placement strategies. However, the unique features of PASS---such as near-field effects, LoS-dominated links, and location-flexible pinching---necessitate the development of more accurate spatial models to capture the spatial randomness, which can be a promising and interesting research direction. Furthermore, although PASS system is LoS-dominated, realistic models covering LoS blockage and residual multipath are still essential for practical performance evaluation.
    \item \textbf{Machine Learning Empowered PASS}: Machine learning empowers both channel estimation and efficient pinching deployment for PASS. Owing to the large spatial extent, PASS may observe non-stationary multi-path characteristics in different serving areas. Generative learning can capture the spatial non-stationarity and reconstruct accurate PASS channels. Moreover, 
    low-complexity learning-based designs can handle the nonconvex and coupled optimization even for a large number of PAs, thus manipulating PAs for both path loss compensation and constructive interference.
    \item \textbf{Advanced Hardware Designs for PASS}: Pinching beamforming plays a critical role in determining the flexibility of PASS and the maximum achievable performance gain. Advanced PASS hardware designs are essential in this context, posing several challenges, such as real-time and continuous control of PA movements or activation, active control of power leakage from PAs, mitigation of unwanted power losses, and desired dielectric materials. Furthermore, hardware intended for wideband transmission must also consider waveguide dispersion, cut-off frequency, and the frequency-dependent coupling factor. These require multi-disciplinary research efforts. 
\end{itemize}

\vspace{-0.1cm}
\bibliographystyle{IEEEtran}
\bibliography{mybib}

\vspace{-0.1cm}
\section*{Biographies}
\vskip -2.5\baselineskip plus -1fil

\begin{IEEEbiographynophoto} {Yuanwei Liu} (Fellow, IEEE) is a Professor at The University of Hong Kong.
\end{IEEEbiographynophoto}
\vskip -2.5\baselineskip plus -1fil
\begin{IEEEbiographynophoto} {Zhaolin Wang} (Member, IEEE) is a Research Assistant Professor at The University of Hong Kong.
\end{IEEEbiographynophoto}
\vskip -2.5\baselineskip plus -1fil
\begin{IEEEbiographynophoto} {Xidong Mu} (Member, IEEE) is a Lecturer with the Centre for Wireless Innovation (CWI), Queen’s University Belfast.
\end{IEEEbiographynophoto}
\vskip -2.5\baselineskip plus -1fil
\begin{IEEEbiographynophoto} {Chongjun Ouyang} (Member, IEEE) is a Post-Doctoral Researcher at Queen Mary University of London. 
\end{IEEEbiographynophoto}
\vskip -2.5\baselineskip plus -1fil
\begin{IEEEbiographynophoto} {Xiaoxia Xu} (Member, IEEE) is a Post-Doctoral Researcher at Queen Mary University of London.
\end{IEEEbiographynophoto}
\vskip -2.5\baselineskip plus -1fil
\begin{IEEEbiographynophoto} {Zhiguo Ding} (Fellow, IEEE) is a Professor in Communications at The University of Manchester and a Distinguished Adjunct Professor at Khalifa University.
\end{IEEEbiographynophoto}

\end{document}